\documentstyle[epsf]{preprint}
\journalid{TSUJIMOTO et al.}{The $r$-process in the Early Galaxy}{2000}
\articleid{1}{6}

\def\Msun{M_{\odot} }
\def\cm3{{\rm ~cm}^{-3}}
\def\ltsima{$\; \buildrel < \over \sim\;$}
\def\ltsim{\lower.5ex\hbox{\ltsima}}
\def\gtsima{$\; \buildrel > \over\sim \;$}
\def\gtsim{\lower.5ex\hbox{\gtsima}}
\def\ms{$M_{\odot}$ }

\def\apj{ApJ}
\def\aj{AJ}

\title{PROBING THE SITE FOR r-PROCESS NUCLEOSYNTHESIS WITH ABUNDANCES 
OF BARIUM AND MAGNESIUM IN EXTREMELY METAL-POOR STARS}{The $r$-process 
in the Early Galaxy}
\author{Takuji Tsujimoto \\
{\em National
Astronomical Observatory, Mitaka-shi, Tokyo, 181-8588 Japan;} \\ 
{\em taku.tsujimoto@nao.ac.jp} \\
Toshikazu Shigeyama$^{1}$ and Yuzuru Yoshii$^{1,\,2}$ \\
1) {\em Research Center for the Early Universe, Graduate School of 
Science, University of Tokyo,} \\ {\em Bunkyo-ku, Tokyo, 113-0033 Japan} \\
2) {\em Institute of Astronomy, Graduate School of Science, University of 
Tokyo, Mitaka-shi, Tokyo, 181-8588 Japan}}{T. Tsujimoto et al.}

\received{26 October 1999}
\accepted{12 January 2000}
\abstract{
We suggest that if the astrophysical site for $r$-process
nucleosynthesis in the early Galaxy is confined to a narrow mass range
of Type II supernova (SN II) progenitors, with a lower mass limit of
$M_{\rm ms}=20\Msun$, a unique feature in the observed distribution of
[Ba/Mg] {\it vs.} [Mg/H] for extremely metal-poor stars can be
adequately reproduced.  We associate this feature, a bifurcation of
the observed elemental ratios into two branches in the Mg abundance
interval $-2.7 \le {\rm [Mg/H]} \le -2.3$, with two distinct
processes.  The first branch, which we call the ``$y$''-branch, is
associated with the production of Ba and Mg from individual massive
supernovae.  The derived mass of Ba synthesized in SNe II is
$8.5\times 10^{-6}\Msun$ for $M_{\rm ms}=20\Msun$, and $4.5\times
10^{-8}\Msun$ for $M_{\rm ms}=25\Msun$.  We conclude that SNe II with
$M_{\rm ms}\approx 20\Msun$ are the dominant source of $r$-process
nucleosynthesis in the early Galaxy.  An SN-induced chemical evolution
model with this $M_{\rm ms}$-dependent Ba yield creates the
$y$-branch, reflecting the different nucleosynthesis yields of [Ba/Mg]
for each SN II with $M_{\rm ms}\gtsim20\Msun$.  The second branch,
which we call the ``$i$''-branch, is associated with the elemental
abundance ratios of stars which were formed in the dense shells of the
interstellar medium swept up by SNe II with $M_{\rm ms}<20\Msun$ that
do {\it not } synthesize $r$-process elements, and applies to stars
with observed Mg abundances in the range [Mg/H]$ < -2.7$.  The Ba
abundances in these stars reflect those of the interstellar gas at the
(later) time of their formation.
The existence of a [Ba/Mg] $i$-branch strongly suggests that SNe II
which are associated with stars of progenitor mass $M_{\rm ms}\le
20\Msun$ are infertile sources for the production of $r$-process
elements.  We predict the existence of this $i$-branch for other
$r$-process elements, such as europium (Eu), to the extent that their
production site is in common with Ba.}


\begin{document}
\section{INTRODUCTION}

Truran (1981) suggested that the observed presence of elements heavier
than iron (Fe) in extremely metal-poor stars is due to the
nucleosynthesis products of rapid neutron capture reactions (the
$r$-process).  Europium (Eu), for which about 97\% of its abundance in
the solar system is of pure $r$-process origin (K\"appler, Beer, \&
Wisshak 1989), has therefore been considered as one of the most useful
elements for locating the astrophysical sites of $r$-process
nucleosynthesis.  Assuming that such sites would most likely be
identified with Type II supernovae (SNe II), many authors have
attempted to constrain the applicable mass range of SN II progenitors
from an observed [Eu/Fe] {\it vs.} [Fe/H] relation for metal-poor
stars.  However, a satisfactory mass range for the progenitor stars
has not yet been agreed upon.  Mathews, Bazan, \& Cowan (1992)
suggested a mass range of $M_{\rm ms}=7-8\Msun$ for the astrophysical
$r$-process site, while Travaglio et al.~(1999) supported a somewhat
higher mass range, $M_{\rm ms}=8-10\Msun$.  Most recently, Ishimaru \&
Wanajo (1999) concluded that the [Eu/Fe] versus [Fe/H] relation alone
is unable to distinguish an $r$-process site with progenitor masses
$M_{\rm ms}=8-10\Msun$ from that resulting from a site with $M_{\rm
ms}>30\Msun$.

The discovery of numerous Galactic halo stars with extremely low metal
abundances from the ongoing HK survey of Beers and colleagues (Beers,
Preston, \& Shectman 1992; Beers 1999), in particular those in the
abundance range $-4.0 \le {\rm [Fe/H]} \le -2.5$, have opened the door
for much more detailed investigations of the elemental production from
SNe II in the early Galaxy.  Shigeyama \& Tsujimoto (1998) recently
argued that the elemental abundance patterns observed in the
atmospheres of extremely metal-poor stars retain those produced by
{\it individual} SNe II (see also Audouze \& Silk 1995), and thus can
be used to estimate the heavy element yields of the first generation
of stars in the Galaxy.  They also argued, on the basis of theoretical
SN progenitor models, that the synthesized mass of magnesium, $M_{\rm
Mg}$, as a function of $M_{\rm ms}$, is likely to be far more certain
than the predicted mass of iron produced, $M_{\rm Fe}$, which is
dependent on the mass cut chosen in the stellar core.  These
considerations indicate that the observed [Eu/Mg] {\it vs.} [Mg/H]
relation in extremely metal-poor stars should give a much firmer
constraint on the $r$-process site. According to Tsujimoto \&
Shigeyama (1998), stars with the lowest value of [Mg/H]$=-2.7$ (CS
22892-052) that exhibit absorption lines of Eu in their spectra in the
sample of McWilliam et al.~(1995) set the lower limit to $M_{\rm
ms}\approx 20\Msun$, above which SNe II produce the $r$-process
elements.

\begin{figure}[ht]
\begin{center}
\leavevmode
\epsfxsize=\columnwidth\epsfbox{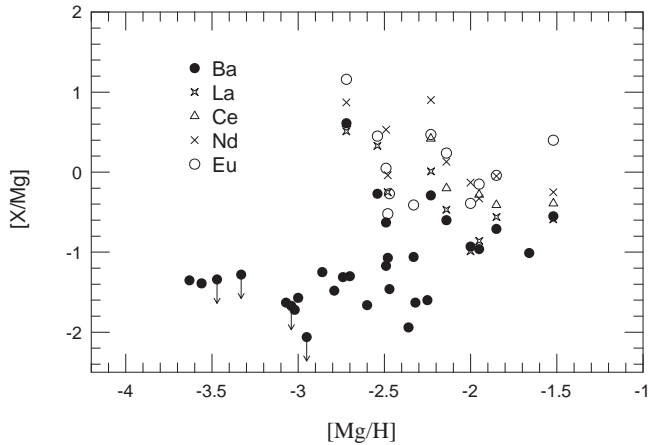}\hfil
\end{center}
\caption{The correlation of [Ba/Mg]
with [Mg/H] for metal-poor stars.  Filled circles show the data of
McWilliam (1998).  Two stars (CS 22898$-$027, CS 22947$-$187) in his
sample are excluded from the plots because their Ba abundances may be
contaminated by $s$-process material accreted from binary companions
(McWilliam 1998).  Other $r$-process data (McWilliam et al.~1995) are
also shown by different symbols indicated in the figure.}
\end{figure}

For most stars of extremely low metallicity ([Fe/H] \ltsim $-$ 2.0),
obtaining an abundance measurement of Eu with 2.5--4m-class telescopes
is a difficult observational task.  For example, McWilliam et
al.~(1995) only obtained Eu abundances for 11 of the 33 metal-poor
stars in their sample.  This incompleteness implies that one might
miss valuable insights to $r$-process production mechanisms by
concentrating exclusively on observations of Eu.  McWilliam (1998)
showed that the [Ba/Eu] ratios for stars with $-3\ltsim$[Fe/H]$\ltsim
-2$ are consistent with pure $r$-process ratios, and concluded that Ba
in extremely metal-poor stars is also of $r$-process origin.  Burris
et al.~(1999) arrived at a similar result.

In Figure 1, which uses the data of McWilliam et al.~(1995) and
McWilliam (1998), the [Ba/Mg] values for a sample of metal-poor stars
are plotted against [Mg/H], together with four heavier $r$-process
elements, La, Ce, Nd, and Eu.  It is indeed remarkable that there
appears a nearly vertical boundary at [Mg/H]$\sim -2.5$, with [Ba/Mg]
spanning the range from [Ba/Mg]$=-2$ to +0.6, whereas a horizontal
line of [Ba/Mg]$\sim -1.4$ emerges from [Mg/H]$\approx -2.5$ down to
$-3.7$.  The data of Ryan, Norris, \& Beers (1996) for seven of their
extremely metal-poor stars show a similar trend in the
[Ba/Mg]$-$[Mg/H] plane.

In \S 2 we discuss the $r$-process sites which might be consistent
with the vertical [Ba/Mg]$-$[Mg/H] boundary, based on the models
proposed by Shigeyama \& Tsujimoto (1998) and Tsujimoto \& Shigeyama
(1998). In \S 3 the entire range of observed [Ba/Mg]$-$[Mg/H]
abundances in halo stars is used to explore the early stages of
inhomogeneous chemical evolution in the Galaxy in the context of the
SN-induced star-formation model described by Tsujimoto, Shigeyama, \&
Yoshii (1999, hereafter TSY).

\section{PRODUCTION SITE FOR r-PROCESS ELEMENTS}

Figure 2a illustrates the hypothesis we seek to defend, that the
metal-poor stars of the Galaxy populate two separate branches in the
[Ba/Mg]$-$[Mg/H] plane.  The first branch extends rightward of the
vertical boundary which begins at ([Ba/Mg], [Mg/H])=($-2.0, -2.3$) and
ends at ([Ba/Mg], [Mg/H])=($+0.6, -2.7$).  The second branch is
horizontal from [Mg/H]$\sim -2.5$ to $-3.7$ at a constant value of
[Ba/Mg]$\sim -1.4$.

\begin{figure}[ht]
\begin{center}
\leavevmode
\epsfxsize=\columnwidth\epsfbox{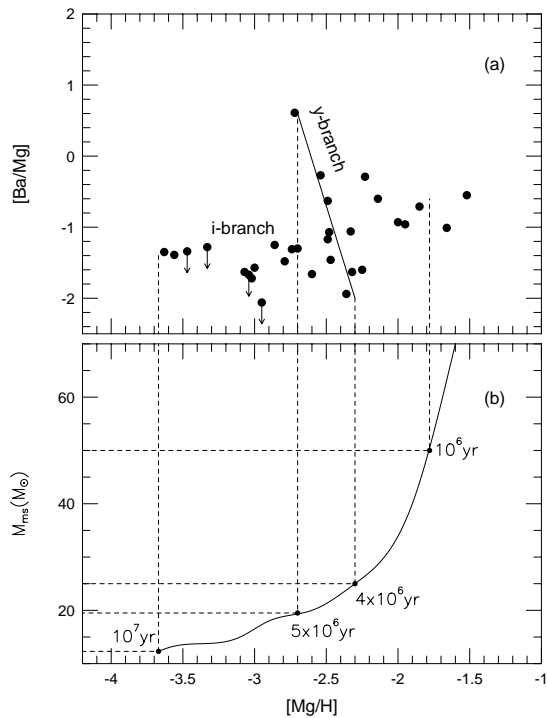}\hfil
\end{center}
\caption{($a$) The correlation of [Ba/Mg] with [Mg/H] for metal-poor 
stars; same as in Fig.1. The thick line is the left-side boundary to
the $y$-branch representing the mass range of the $r$-process site.
($b$) SN II progenitor mass $M_{\rm ms}$ plotted as a function of
[Mg/H] inside the shell swept up by the SNR.  The stellar lifetimes
are also shown for four reference masses of $M_{\rm ms}=12$, 20, 25,
and $50\Msun$.  Dashed lines in the lower and upper panels illustrate
how we assign $M_{\rm ms}$ to each star along the boundary.}
\end{figure}

Other $r$-process elements, such as La, Ce, Nd, and Eu, shown in
Figure 1, populate the first branch identified from the behavior of
[Ba/Mg] in Figure 2, but we presently lack measurements of their
abundances in the most metal-poor stars, which are required in order
to confirm the presence of the second branch.  Tsujimoto \& Shigeyama
(1998) derived the Eu yield as a function of SN II progenitor mass,
and showed that average predicted value, weighted by the Salpeter
initial mass function (IMF), successfully reproduces a plateau value
of [Eu/Fe] seen at $-2\ltsim$[Fe/H]$\ltsim -1$.  Thus, hereafter, we
refer to the first branch as the $y$-branch, where the letter ``$y$''
stands for its origination in individual SNe II yields.

As shown in Figure 2a, the $y$-branch is confined to stars with
[Mg/H]$>-2.7$; there is no star with [Mg/H]\ltsim$-2.7$ that belongs 
to the $y$-branch. On the other hand, Shigeyama \& Tsujimoto (1998)
showed that the metallicity [Mg/H] of stars born from an SN remnant
(SNR) is well approximated by the average [Mg/H] inside the shell
swept up by the SNR. This gives a relation between the metallicity
[Mg/H] of stars and the mass $M_{\rm ms}$ of SN II progenitor as shown
in Figure 2b.  As a consequence, it is indicated that only SNe II with
$M_{\rm ms}$\gtsim $20\Msun$ produce Ba via the $r$-process. If we
assume that the vertical boundary to the $y$-branch has a one-to-one
correspondence to the Ba yield from individual SNe II, the progenitor
mass range is confined to $M_{\rm ms}=20-25\Msun$.  It is
straightforward to derive the Ba yield from the observed [Ba/Mg]-value
along the vertical boundary and the synthesized Mg mass, leading to
$M_{\rm Ba}=8.5\times 10^{-6}\Msun$ for $M_{\rm ms}=20\Msun$, and
$M_{\rm Ba}=4.5\times 10^{-8}\Msun$ for $M_{\rm ms}=25\Msun$.

We now consider the origin of the second branch of [Ba/Mg], in the
range of $-3.7\ltsim$[Mg/H]$\ltsim-2.7$.  This [Mg/H] range
corresponds to $M_{\rm ms}=12-20\Msun$, for which SNe II, according to
our present models, do not produce significant amounts of Ba. 
Therefore, our hypothesis is that the Ba abundances which are observed
in the atmospheres of these stars come only from the interstellar
matter (ISM) that was enriched in Ba by the preceding SNe II (with
$M_{\rm ms}=20-25\Msun$), and that was swept up in the shells by later
SNe II with $M_{\rm ms}=12-20\Msun$. Thus, hereafter we refer to this
branch, as the $i$-branch, where the letter ``$i$'' stands for an ISM
origin. We note that the [Ba/Mg] value in this branch should always be
below its plateau value [Ba/Mg]$\sim -0.6$ at higher metallicities.
We also predict from the above argument that stars born from the shell
swept up by SNe II with $M_{\rm ms}>25\Msun$ would form another
$i$-branch at [Mg/H]$\gtsim -2.5$, evidence for which is not seen in
the sample of McWilliam et al.~(1995).

\section{CHEMICAL EVOLUTION OF $R$-PROCESS ELEMENTS IN THE GALACTIC HALO}

In this section, we discuss the chemical evolution of Ba and Eu in the
Galactic halo, based on the formulation presented in TSY. The essence
of the TSY model is that the star-forming process is assumed to be
confined to separate clouds which make up the entire halo, and that
the chemical evolution in these clouds proceeds through a successive
sequence of SN II explosion, shell formation, and resulting star
formation.  Details can be found in TSY.

An overview of our model applied to the Galactic halo is as follows:
-- (1) The metal-free Population III stars (Pop III) form by some (as
yet unspecified) mechanism in primordial-composition gas clouds of the
Galactic halo.  (2) The most massive stars among them explode as Pop
III SNe II, which trigger a series of star formation events.  (3) Star
formation terminates when SNRs become unable to the
formation of dense shells.  (4) Roughly 90 \% of the cloud mass
remains unused in star formation, and may fall onto the still-forming
Galactic disk.

The free parameters in our model are the mass fraction $x_{\rm III}$
of metal-free Pop III stars initially formed in each cloud, and the
mass fraction $\epsilon$ of stars formed in a dense shell swept up by 
each SNR.  A value of $x_{\rm III}$ sensitively determines the level
of [Ba/Mg] for the $i$-branch, so we take $x_{\rm
III}=2.5\times10^{-4}$ to be consistent with the observed value
[Ba/Mg]=$-1.4$ at [Mg/H]$\sim -3.7$ (McWilliam 1998).  We take
$\epsilon=4.3\times10^{-3}$ to reproduce the observed [Fe/H]
distribution function of halo field stars for [Fe/H]$<-1$ (TSY).

\begin{figure}[ht]
\begin{center}
\leavevmode
\epsfxsize=0.74\columnwidth\epsfbox{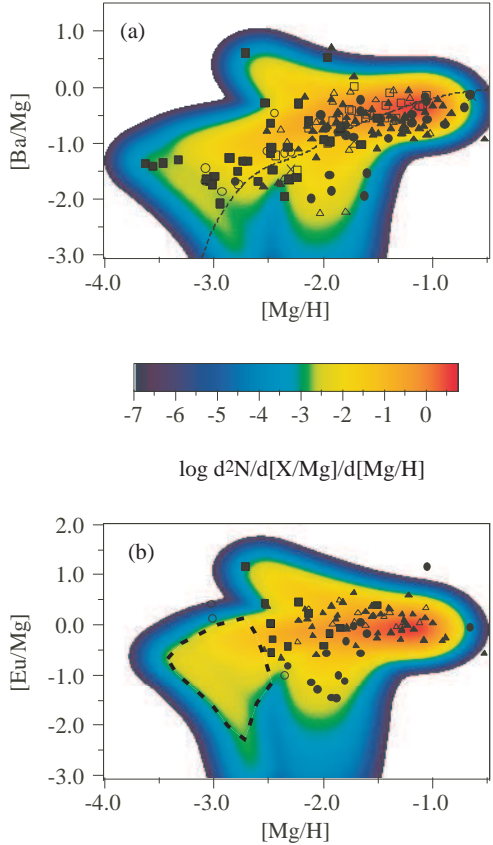}\hfil
\end{center}
\caption{($a$) The color-coded frequency distribution of the
long-lived stars in the [Ba/Mg]$-$[Mg/H] plane, convolved with a
Gaussian having $\sigma$=0.2dex for [Ba/Mg] and $\sigma$=0.1dex for
[Mg/H].  The symbols represent the data taken from McWilliam 1998
({\it filled squares}), Ryan et al.~1996 ({\it open circles}), Burris
et al.~1999 ({\it filled triangles}), Luck \& Bond 1985 ({\it filled
circles}), Luck \& Bond 1981 ({\it open triangles}), Magain 1989 ({\it
open squares}), and Steenbock 1983 ({\it crosses}). For comparison,
the result of the conventional one-zone chemical evolution model is
shown by dashed line assuming the $r$-process site of $8-10$ \ms
supernovae, which is different from our proposed site of $20-25$ \ms
supernovae.  ($b$) The same as ($a$) but for Eu.  The dashed curve
encloses the predicted region in which $i$-branch stars must exist.}
\end{figure}

Adopting the result of \S 2, that SNe II with $M_{\rm ms}=20-25\Msun$
are the dominant site for $r$-process nucleosynthesis, and using the
derived Ba yield dependent on $M_{\rm ms}$, we calculate the Ba
evolution at early epochs of the Galactic halo. For later evolution,
which pertains to the abundance range $-2$\ltsim[Mg/H]\ltsim$-1$, we
add the contribution from $s$-process nucleosynthesis, assuming that
the production site for early-epoch $s$-process elements is $2-3$ \ms
AGB stars (Burris et al.~1999).  The stellar IMF is assumed to be of
the Salpeter form. The upper and lower mass limits for stars that
explode as SNe II are taken to be $50\Msun$ and $12\Msun$,
respectively.

Figure 3a is a color-coded predicted frequency distribution of stars
in the [Ba/Mg]$-$[Mg/H] plane, normalized to unity when integrated
over the entire area (see the color bar for the scale). In order to
enable a direct comparison with the data, the frequency distribution
has been convolved with a Gaussian with $\sigma=0.2$ dex for [Ba/Mg]
and $\sigma=0.1$ dex for [Mg/H]. As already pointed by Burris et
al.~(1999), there is a systematic difference in the derived Ba
abundance between Burris et al.~(1999) and McWilliam (1998). Burris et
al.~(1999) obtain [Ba/H] ratios which are higher, in the mean, by 0.46
dex for the seven stars in common with the data of McWilliam (1998).
We therefore decrease their [Ba/H] ratios by 0.46 dex to enable a
consistent comparison with the data of McWilliam (1998) in Figure 3a. 
The [Ba/Mg] ratio ``plateau'' in the range of
$-2$\ltsim[Mg/H]\ltsim$-1$ is obviously higher than that of $i$-branch
([Mg/H] \ltsim $-2.5$), which implies a significant contribution from
$s$-process nucleosynthesis. For comparison, we show by the dashed
line in this figure the conventional one-zone chemical evolution model
assuming that the production sites for $r$-process and $s$-process are
$8-10$ \ms supernovae and $2-3$ \ms AGB stars, respectively. It is
clear that the conventional model predicts a much smaller [Ba/Mg] than
is consistent with the observed $i$-branch at [Mg/H] \ltsim $-2.7$.

To summarize our hypothesis: The earliest stars, stars that were
formed from the shells swept up by SNe II with $M_{\rm
ms}=20-25\Msun$, first create the vertical boundary seen in Figure 2a.
Thereafter, the $y$-branch develops at higher [Mg/H] and converges to
the plateau level of [Ba/Mg] at higher metallicities, as described in
detail in TSY. At the same time, the subsequently formed stars,
arising from the SNRs with $M_{\rm ms}=12-20\Msun$, do not produce Ba,
and therefore have the lowest value of [Ba/Mg] at [Mg/H]$\ltsim -2.5$. 
These stars populate the lower envelope of the $i$-branch, then fill
in this branch at lower [Mg/H], but gradually {\it increasing}
[Ba/Mg].  As the Mg abundance in the ISM becomes dominant over that
ejected from SNe II, the $i$-branch develops at higher abundances,
increasing both [Mg/H] and [Ba/Mg].

Because the upper mass limit in our present model for the Ba
production site, $25\Msun$, is set below the canonical upper limit
$\sim 50\Msun$ for stable stellar masses, we predict the existence of
another $i$-branch in the range of $-2.3\ltsim$[Mg/H]$\ltsim-1.5$,
which is made up of stars arising from the SNRs with $M_{\rm
ms}=25-50\Msun$.  While this branch does not show up in the sample of
McWilliam (1998), the data from Luck \& Bond (1981,1985),
interestingly, support such a prediction (see filled and open circles
in Fig.3a). If we increase the upper mass limit to $50\Msun$ for the
Ba production site, a small amount of Ba is produced in SNe with
$M_{\rm ms}\gtsim 25\Msun$, which only slightly contaminates the
$i$-branch stars.  The results of our calculations with an upper mass
limit of $M_{\rm ms}=25\Msun$ remain almost unchanged, because of the
assumed declining IMF at high masses.

The value of $x_{\rm III}=2.5\times 10^{-4}$ adopted here can be
converted into a probability of finding Pop III stars, $p_{\rm
III}=2.5\times 10^{-3}$, defined as the expected number of Pop III
stars divided by the total number of the long-lived stars in a sample
(see TSY).  However, we regard this $x_{\rm III}$-value as an upper
bound, because the predicted [Ba/Mg] {\it vs.} [Mg/H] distribution
with smaller $x_{\rm III}$ provides an adequate fit to the observed
distribution.  A much more strict constraint on $p_{\rm III}$ could be
obtained if the sample of extremely metal-poor stars with measured Ba
abundances is increased by a factor of ten.

As a final exercise, we also calculate the Eu evolution in the early
Galaxy using the same models and assumptions we have employed for the
prediction of Ba evolution.  Figure 3b shows the color-coded frequency
distribution of stars in the [Eu/Mg]$-$[Mg/H] plane, after convolving
with $\sigma=0.2$ dex for [Eu/Mg] and $\sigma=0.1$ dex for [Mg/H]. The
Eu production site must be the same as for Ba, or at least share a
part of SNe II sites where Ba is produced, because the rapid
$n$-capture process cannot synthesize Eu without producing Ba. 
Therefore, we assume that Eu is also produced by SNe II with $M_{\rm
ms}=20-25\Msun$. We scale the Ba yield by the pure $r$-process value
[Ba/Eu]=$-0.72$ (Wisshak, Voss, \& K\"apeler 1996) and obtain a
predicted Eu yield of $M_{\rm Eu}=1.3\times 10^{-6}\Msun$ for $M_{\rm
ms}=20\Msun$, and $M_{\rm Eu}=7.0\times 10^{-9}\Msun$ for $M_{\rm
ms}=25\Msun$.  Our calculation predicts the existence of a Eu
$i$-branch at [Mg/H]$\ltsim -2.5$, which must be confirmed by future
measurements of Eu abundances for extremely metal-poor
stars.{\footnote[2]{ In TSY, we used the data of Ryan et al.~(1996)
for determining the Eu yield.  These observations exhibit a much lower
[Eu/Mg] ratio than that of McWilliam et al.~(1995). As a consequence,
the entire population of SNe II with $M_{\rm ms}=12-50\Msun$ may be
the Eu production site.}

\section{CONCLUSION}

We have suggested the existence of two distinct correlations of
[Ba/Mg] with [Mg/H] for metal-poor stars in the Galaxy, the $y$- and
$i$-branches, separated at [Mg/H] $\sim -2.7$. These branches cross
one another almost perpendicularly, and form an arrow-like frequency
distribution of stars in the [Ba/Mg]$-$[Mg/H] plane.  The vertical
boundary to the $y$-branch extending from [Ba/Mg]$=-2.0$ to $+0.6$
reflects the different nucleosynthesis contributions of [Ba/Mg] for
each SN II of progenitor mass in a narrow range, $M_{\rm
ms}=20-25\Msun$.  The horizontal $i$-branch extending from
[Mg/H]$=-2.7$ to $-3.7$ is populated by stars born from the shells
swept up by SNe II with $M_{\rm ms}=12-20\Msun$ that do not synthesize
the $r$-process elements.  Therefore, the Ba abundances of the
$i$-branch stars reflect those in the ISM at the time when such stars
form.

The $i$-branch provides strong evidence for the existence of a lower
mass limit of the SN progenitors for the site of $r$-process
nucleosynthesis.  The upper mass limit for the $r$-process site cannot
be determined well from the current observations.  Because of the
declining IMF, SNe II with $M_{\rm ms}\gtsim 25\Msun$ contributes
little to the total Ba production in the early Galaxy. In other words,
in our formulation, the $r$-process site is dominated by SNe II with
$M_{\rm ms}\approx 20-25\Msun$.

The $i$-branch should not exist for $\alpha$- and Fe-peak heavy
elements formed in the early Galaxy, because the entire range of SNe
II progenitor masses will contribute to substantial amounts of such
elements.  However, the $i$-branch is expected to exist, though is not
confirmed as of yet, in the case of the $r$-process elements such as
La, Ce, Nd, and Eu. It is therefore important to measure the
abundances of such elements for the stars CS 22878$-$101, CS
22897$-$008, CS 22891$-$200, CS 22891$-$209, CS 22948$-$066, CS
22952$-$015, and CS 22950$-$046 in the sample of McWilliam et
al.~(1995), which actually form the Ba $i$-branch.

\bigskip

We are grateful to Timothy C. Beers for his careful reading and
helpful comments on our Letter. This work has been partially supported
by COE research (07CE2002) and a Grant-in-Aid for Scientific Research
(11640229) of the Ministry of Education, Science, Culture, and Sports
in Japan.


\begin{thebibliography}{}
\bibitem[]{}
Audouze, J., \& Silk, J. 1995, \apj, 451, L49
\bibitem[]{}
Beers, T. C., in The Third Stromlo Symposium: The Galactic Halo, eds. 
B. Gibson, T. Axelrod, \& M. Putman (ASP: San Francisco), 165, 206
\bibitem[]{}
Beers, T. C., Preston, G. W., \& Shectman, S. A. 1992, \aj, 103, 1987
\bibitem[]{}
Burris, D. L., Pilachowski, C. A., Armandroff, T. E., Sneden, C., 
Cowan, J. J., \& Roe, H. 1999, \apj, in press
\bibitem[]{}
Ishimaru, Y., \& Wanajo, S. 1999, \apj, 511, L33
\bibitem[]{}
K\"appler, F., Beer, H., \& Wisshak, K. 1989, Rep.Prog.Phys., 52, 945
\bibitem[]{}
Luck, R. E., \& Bond, H. E. 1981, {\apj}, 244, 919
\bibitem[]{}
Luck, R. E., \& Bond, H. E. 1985, {\apj}, 292, 559
\bibitem[]{}
Magain, P. 1989, A\&A, 209, 211
\bibitem[]{}
Mathews, G. J., Bazan, G., \& Cowan, J. J. 1992, \apj, 391, 719
\bibitem[McWilliam, Preston, Sneden, \& Searle 1995]{mcwilliam95}
McWilliam, A., Preston, G. W., Sneden, C., \& Searle, L. 1995, \aj, 109, 2757
\bibitem[McWilliam 1998]{mcwilliam98}
McWilliam, A. 1998, \aj, 115, 1640
\bibitem[Ryan, Norris, \& Beers 1996]{ryan96}
Ryan, S. G., Norris, J. E., \& Beers, T. C. 1996, \apj, 471, 254
\bibitem[Shigeyama \& Tsujimoto 1998]{shigeyama98}
Shigeyama, T., \& Tsujimoto, T. 1998, \apj, 507, L135
\bibitem[Travaglio et al., 1999]{travaglio99}
Steenbock, W. 1983, A\&A, 126, 325
\bibitem[Travaglio et al., 1999]{travaglio99}
Travaglio, C., Galli, D., Gallino, R., Busso, M., Ferrini, F., \&
Straniero, O. 1999, \apj, 521, 691
\bibitem[Tsujimoto \& Shigeyama 1998]{tsujimoto98}
Tsujimoto, T., \& Shigeyama, T. 1998, \apj, 508, L151
\bibitem[Tsujimoto, Shigeyama, \& Yoshii 1999]{tsujimoto99}
Tsujimoto,
T., Shigeyama, T., \& Yoshii, Y. 1999, \apj, 519, L64
\bibitem[Wisshak,
Voss, K\"appeler 1996]{Wisshak96}
Wisshak, K., Voss, F., \& K\"apeler, F. 1996 in Proc. 8th Workshop on Nuclear
Astrophysics, ed. W. Hillebrandt \& E. M\"uller (Garching: MPIA),
16
\end{thebibliography}
\end{document}